\documentclass{iau_FM}
\usepackage{graphicx}

\newcommand{\pdf}{PDF($j/j_{mean}$)} 

\title[Spatially-resolved AM] 
{Spatially-resolved galaxy angular momentum}

\author[Sarah M. Sweet et al.]   
{Sarah M. Sweet$^1$,
Deanne B. Fisher$^1$,
Karl Glazebrook$^1$,
Danail Obreschkow$^2$,
Claudia D. P. Lagos$^2$,
 \and Liang Wang$^2$}

\affiliation{$^1$Centre for Astrophysics and Supercomputing, Swinburne University of Technology, PO Box 218, Hawthorn, VIC 3122, Australia \\ email: {\tt sarah@sarahsweet.com.au} \\[\affilskip]
$^2$International Centre for Radio Astronomy Research, University of Western Australia, 7 Fairway, Crawley, WA 6009, Australia}

\pubyear{2018}
\jname{Astronomy in Focus, Volume 1} 
\editors{Teresa Lago, ed.}
\begin{document}

\maketitle

\begin{abstract}
The total specific angular momentum $j$ of a galaxy disk is matched with that of its dark matter halo, but the distributions are different, in that there is a lack of both low- and high-$j$ baryons with respect to the CDM predictions. I illustrate how \pdf\/ can inform us of a galaxy's morphology and evolutionary history with a spanning set of examples from present-day galaxies and a galaxy at $z\sim 1.5$. The shape of \pdf\/ is correlated with photometric morphology, with disk-dominated galaxies having more symmetric \pdf\/ and bulge-dominated galaxies having a strongly-skewed \pdf\/. Galaxies with bigger bulges have more strongly-tailed \pdf\/, but disks of all sizes have a similar \pdf\/. In future, \pdf\/ will be useful as a kinematic decomposition tool.
\end{abstract}

\firstsection 
\section{Introduction}

Angular momentum (AM) is a fundamental parameter in the evolution of galaxies. A dark matter (DM) halo spinning up in the early universe is subject to the same tidal torques as the baryons at its centre, so the total AM of both components is linked, and the specific AM, $j = J/M$ of the baryons is well matched to $j$ of the DM, (e.g \cite[Catalan \& Theuns 1996]{Catalan+1996}). 
$j$ is connected to photometric morphology via the stellar mass -- specific AM -- morphology plane, first shown by \cite[Fall 1983]{Fall83}, such that galaxies with higher $M_*$ have higher $j$, modulo morphology, with the relation for earlier type galaxies offset to lower $j$. This has since also been shown by \cite[Romanowsky \& Fall (2012)]{RF12}, \cite[Obreschkow \& Glazebrook (2014)]{OG14}, \cite[Cortese et al. (2016)]{Cortese+2016}, \cite[Posti et al. (2018)]{Posti+2018}, and \cite[Sweet et al. (2018)]{Sweet+2018}.

\begin{figure}
\begin{center}
 \includegraphics[width=0.4\linewidth]{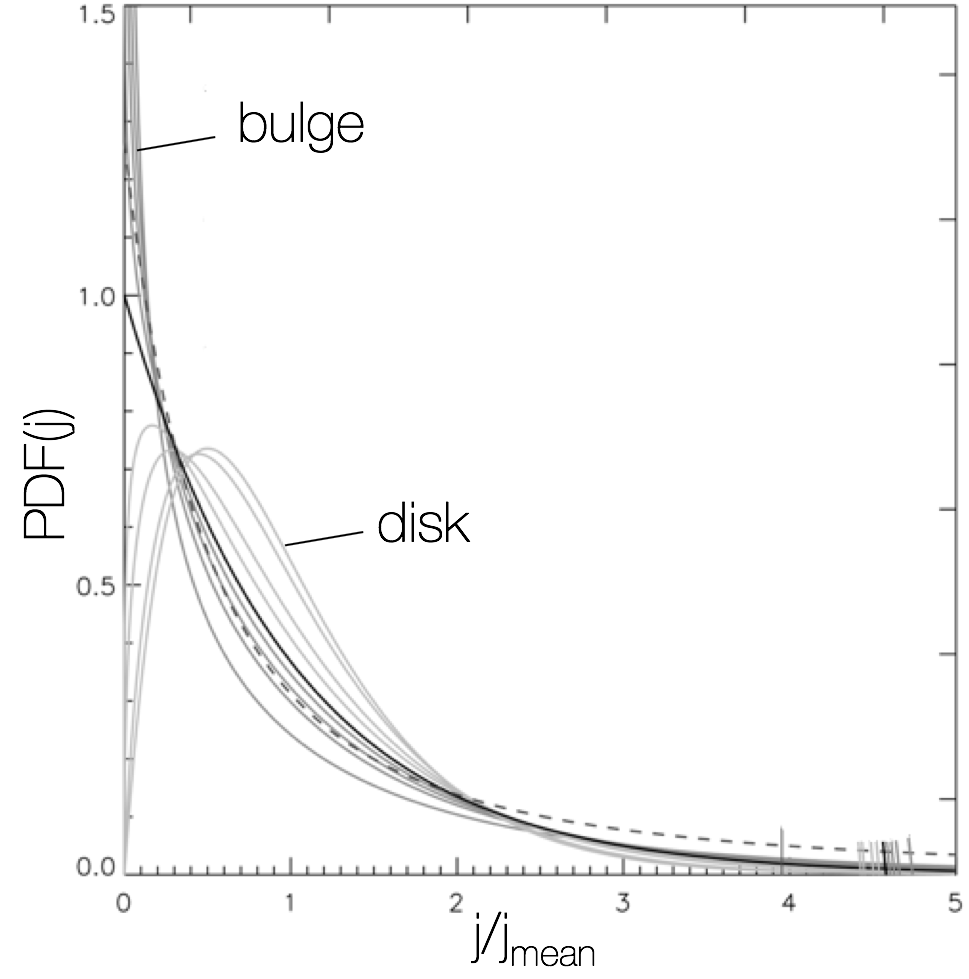} 
 \caption{Predictions from \cite[Sharma \& Steinmetz 2005]{SS05} for baryonic galaxy components. The \pdf\/ for bulge peaks at $j=0$, while disk components have an exponential profile.}
   \label{ss05}
\end{center}
\end{figure}

Although the total $j$ for baryons and DM is linked, further physical processes affect the distribution of $j$ for baryons. \cite[van den Bosch et al. (2001)]{vdb+01} studied the probability density function of $j$ normalised to the mean of the galaxy, \pdf, and found that dwarf galaxies had a deficit of high-$j$ and of low-$j$ material with respect to the prediction for a DM halo. They attributed this to tidal stripping of the outer, rapidly-rotating material and feedback ejecting the inner, dispersion-dominated material respectively. \cite[Sharma \& Steinmetz (2005)]{SS05} then predicted the \pdf\/ for baryonic components (see Fig.\,\ref{ss05}), where bulges, which are dominated by random motions, exhibit a peak at $j=0$, and disks have a \pdf\/ of the form $x$exp$(-kx)$ due to their well-ordered rotation. Also see our updated predictions using the NIHAO simulations (Wang+ in prep)).

The \pdf\/ encodes more physical information than photometry alone, so in this work I am investigating the utility of \pdf\/ as a kinematic tracer of morphology and kinematic decomposition tool.

\section{\pdf\/}

I have constructed \pdf\/ for a subset of the Calar Alto Legacy Integral Field Area survey (CALIFA, \cite[Sanchez et al. 2012]{Sanchez+2012}), using observations of 25 galaxies where the stellar kinematics reach to three times the effective radius. I calculate $j_i = r_i \times v_i$ in every spaxel $i$, where the velocity $v_i = sqrt( v_{i,circ}^2 + v_{i,disp}^2)$ includes the circular velocity $v_{i,circ}$ and dispersion $v_{i,disp}$ added in quadrature. The map of $j$ is then weighted by stellar surface density, as a proxy for mass, and the histogram plotted as the \pdf\/. 

\begin{figure}
\begin{center}
 \includegraphics[width=0.495\linewidth]{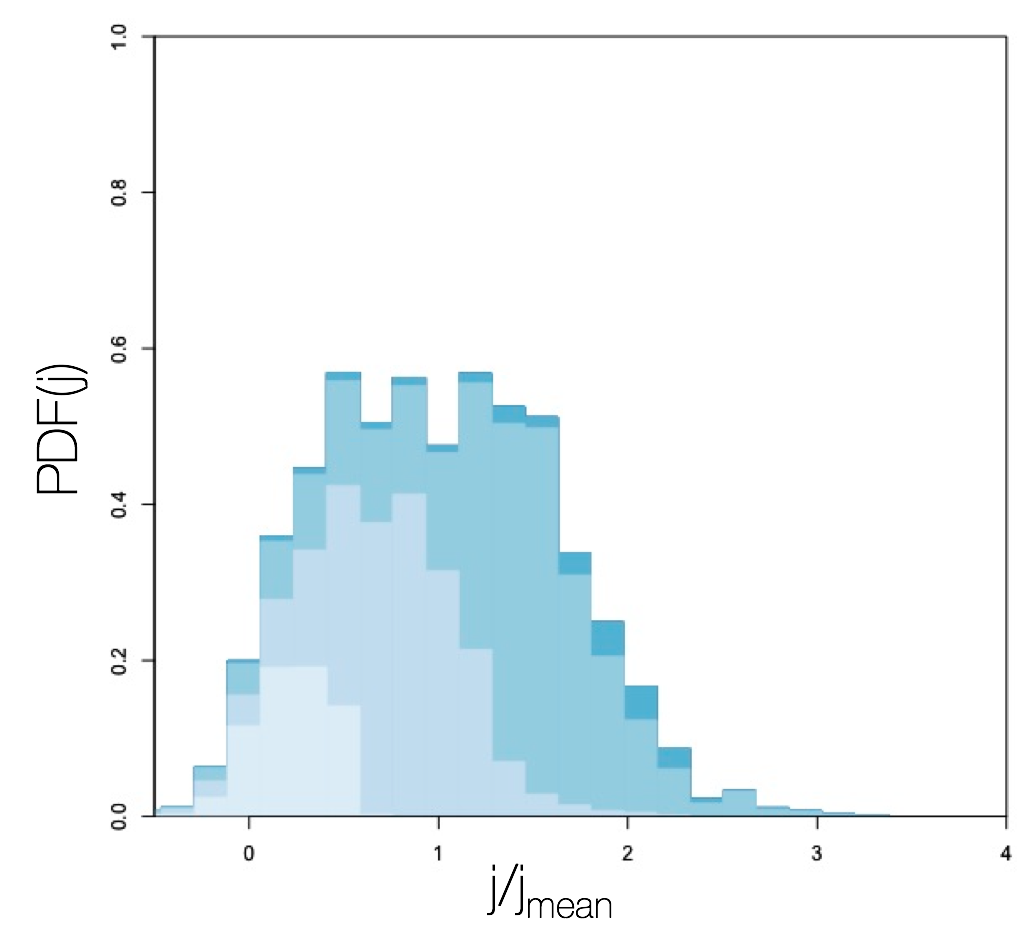} 
 \includegraphics[width=0.495\linewidth]{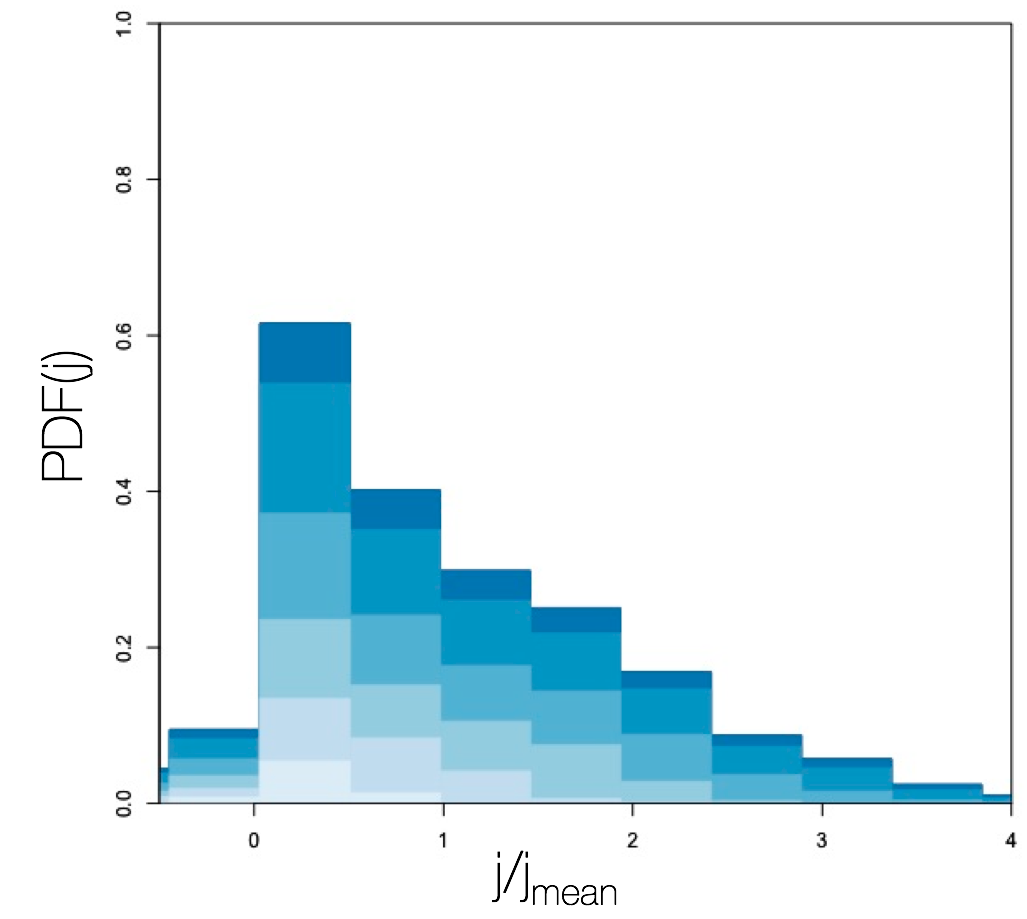} 
 \includegraphics[width=0.495\linewidth]{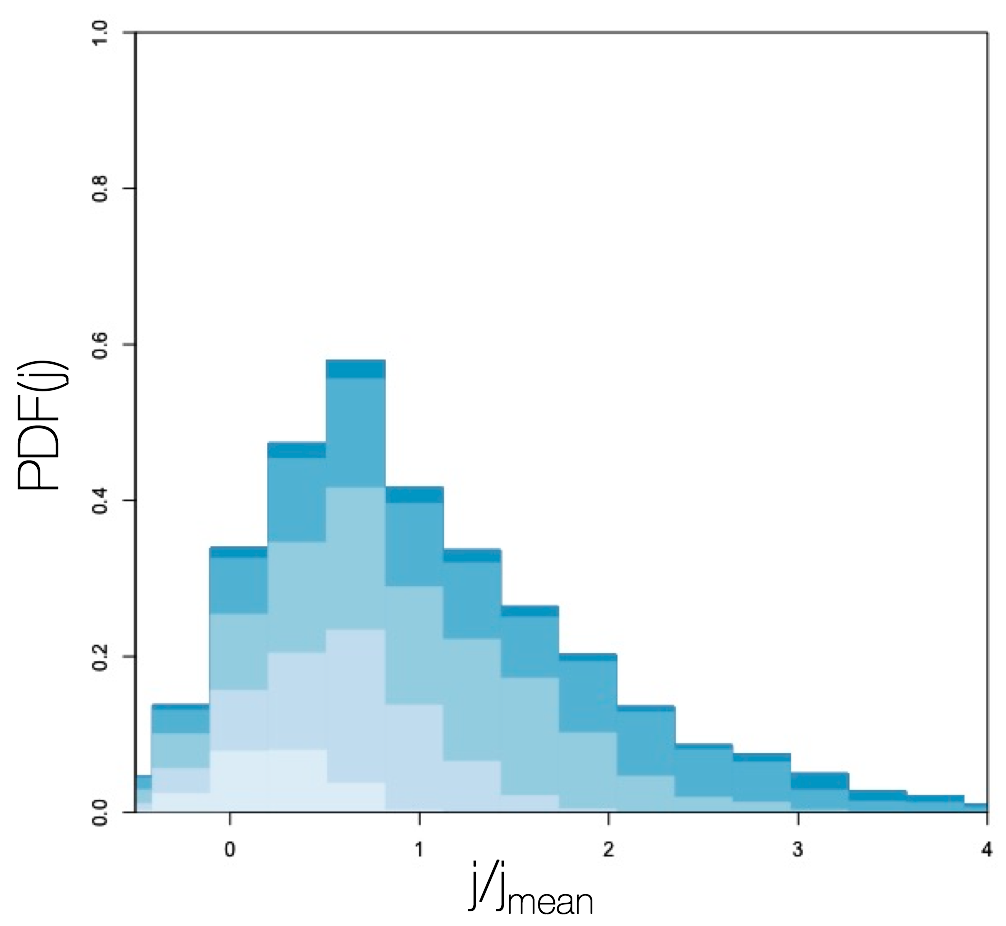} 
 \includegraphics[width=0.495\linewidth]{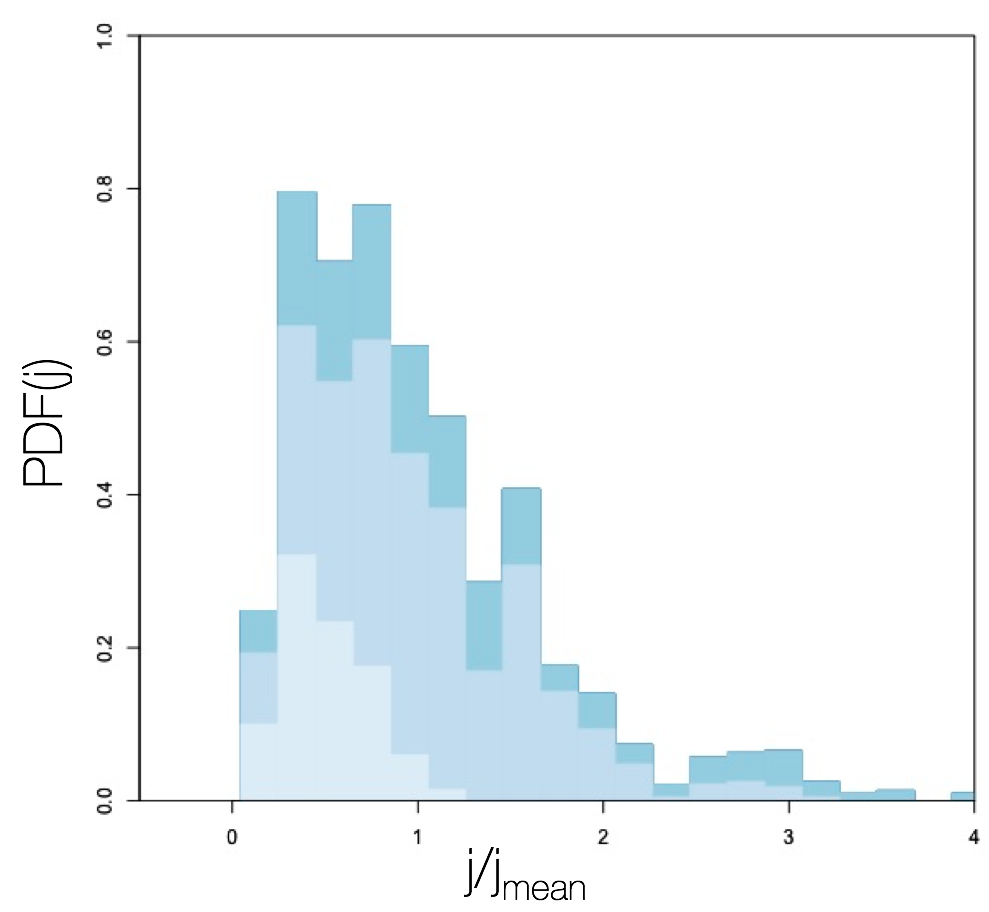} 
 \caption{Example \pdf\/ for local galaxies in the CALIFA sample and one disk at $z\sim 1.5$. Top left: late-type spiral NGC 6063 with low B/T ratio has a broad, symmetric \pdf\/ which peaks near 1. Top right: early type NGC 2592 with high B/T ratio has a strongly-skewed \pdf\/ which peaks nearer 0. Bottom left: NGC 7653 with moderate B/T ratio has a \pdf\/ which is intermediate between the two extremes. Bottom right: $z\sim 1.5$ clumpy disk galaxy COSMOS 127977 also has an intermediate \pdf\/.}
   \label{examples}
\end{center}
\end{figure}

A spanning set of local examples is shown in Fig.\,\ref{examples}. The colour represents radial distance, with lighter colours indicating material nearer the centre. NGC 6063 is a late-type spiral galaxy with low bulge-to-total light ratio B/T = 0.04. Its \pdf\/ is broad and symmetric, and peaks near 1, reminiscent of the predictions by  \cite[Sharma \& Steinmetz (2005)]{SS05} for pure disks. NGC 2592 is an early type galaxy with large B/T = 0.54; its \pdf\/ is strongly-skewed and peaks near $j=0$ like the spheroidal components in \cite[Sharma \& Steinmetz (2005)]{SS05}. Intermediate between these two extremes, the \pdf\/ for NGC 7653 (B/T = 0.33) has characteristics of both disk and bulge. Unfortunately the spatial resolution, which dictates the number of bins, is not sufficient to resolve separate components in the \pdf\/.

I also show a clumpy disk galaxy at $z\sim 1.5$, using a combination of OSIRIS adaptive optics integral field spectroscopy to mitigate the effects of beam smearing in the centre of the galaxy, with deeper KMOS seeing-limited data to trace the velocity field out to higher multiples of the effective radius, using a method described in Sweet et al. (in prep.). The shape is intermediate between the pure disk and bulge-dominated local examples, likely due to the typical high-$z$ morphology of dispersion-dominated clumps embedded in a strongly-rotating disk.

\begin{figure}
\begin{center}
 \includegraphics[width=0.45\linewidth]{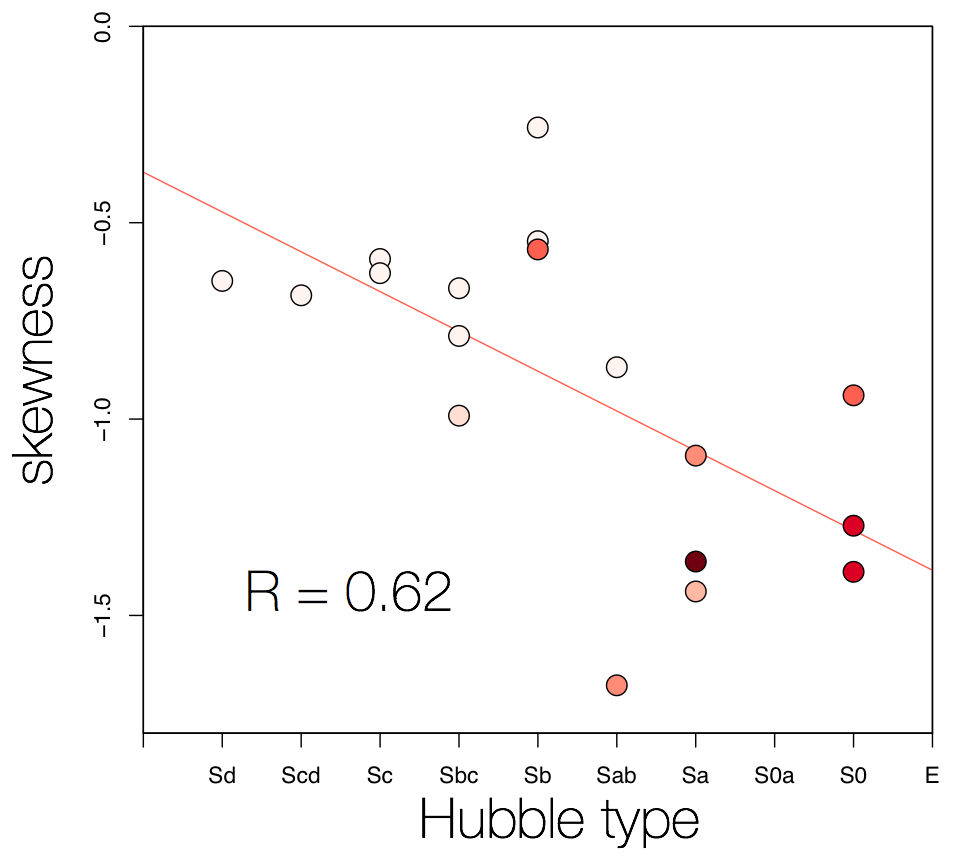} 
 \caption{Correlation between morphology and shape of \pdf\/. Galaxies with earlier Hubble type are more strongly negatively skewed.}
   \label{t-skew}
\end{center}
\end{figure}

There is an apparent trend whereby \pdf that are more skewed and peak nearer $j=0$ correspond to galaxies that are earlier in type and have bigger bulges. This is quantified in the correlation between shape of \pdf\/ (traced by skewness or kurtosis) and morphology (traced by Hubble type and B/T ratio). For instance, Fig.\,\ref{t-skew} illustrates the relation between skewness and Hubble type. Much of the scatter in this correlation arises from the difficulties inherent in photometric classification of Hubble type and quantifying bulge-to-total ratio. Arguably, \pdf\/ as a kinematic quantity encodes more physical information than photometry alone, so may be a more robust tracer of morphology.

\begin{figure}
\begin{center}
 \includegraphics[width=0.8\linewidth]{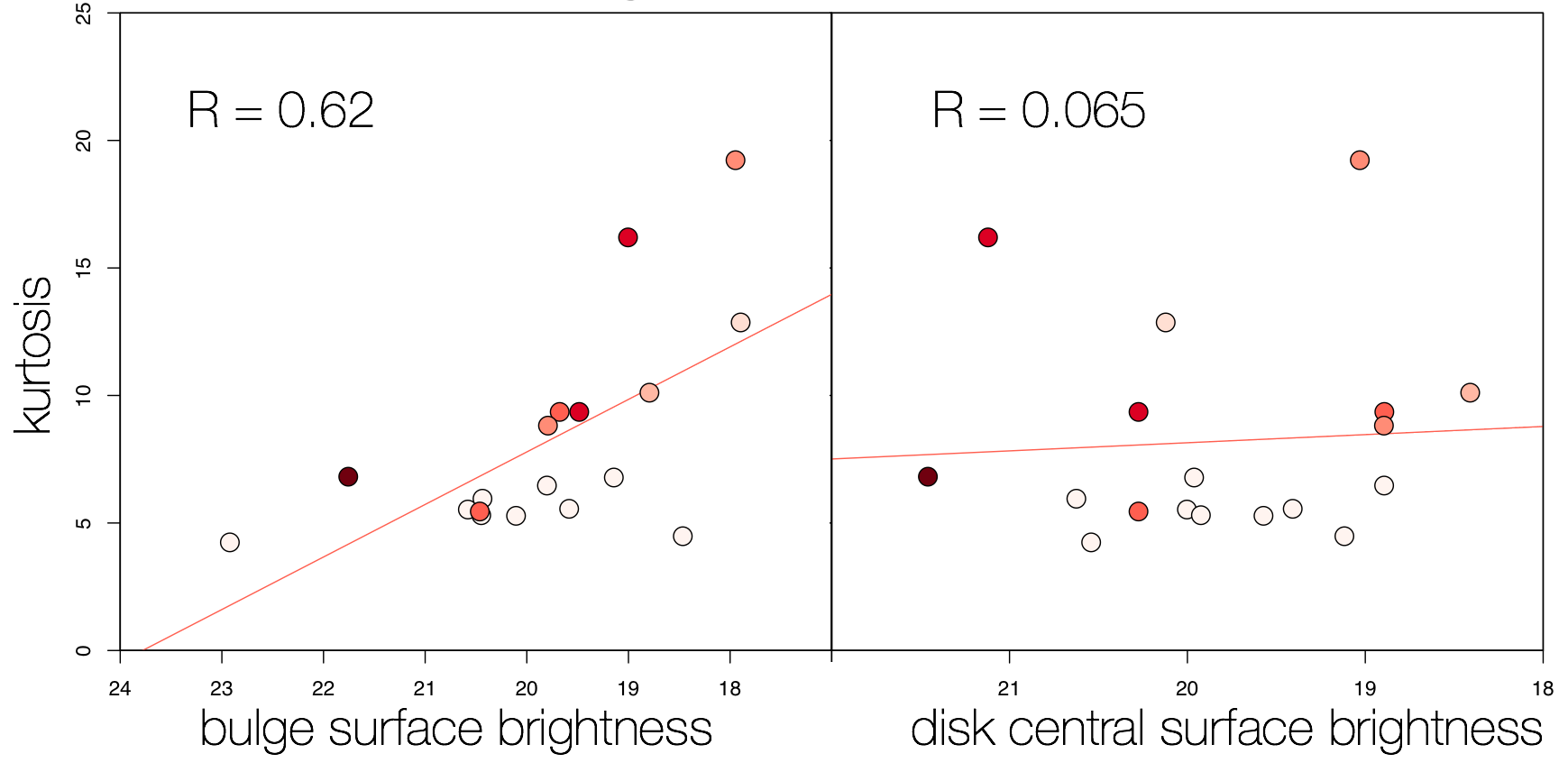} 
 \caption{Correlation between galaxy components and shape of \pdf\/. Galaxies with bigger bulges have more strongly-tailed \pdf\/, but the shape of \pdf\/ is the same for disks of all sizes.}
   \label{bulgediskkurt}
\end{center}
\end{figure}

I also see that the bulge is linked with \pdf\/.  Fig.\,\ref{bulgediskkurt} demonstrates that the shape of \pdf\/ is moderately correlated with the surface brightness of the bulge, such that galaxies with bigger bulges have more strongly-tailed \pdf. On the other hand, the \pdf\/ shape is not at all correlated with the central surface brightness of the disk. This indicates that the size of the bulge is related to the distribution of $j$ within a galaxy, but disks of all sizes have similar \pdf\/.

\section{Conclusion: Utility of PDF(j)}
The \pdf\/ traces kinematic morphology of a galaxy. It encodes more physical information than photometry alone, and is a product of the evolutionary history of the galaxy. In future, as spatial resolution increases, I predict that \pdf\/ will be useful to separate out kinematic components: thin disk from thick disk and bulge, clumps from bulges, and pseudobulges from classical bulges.


%
%
%

\end{document}